\documentclass[a4paper,11pt]{article}
\pdfoutput=1 % if your are submitting a pdflatex (i.e. if you have
\usepackage{jcappub} % for details on the use of the package, please
\usepackage[T1]{fontenc} % if needed
\usepackage{verbatim}
\usepackage{mwe}
\usepackage{subfig}
\usepackage{scrextend}
\usepackage{amsmath}
\usepackage{afterpage}
\usepackage{graphicx} 
\usepackage{color}
\usepackage{lineno}
\usepackage{float}

\newcommand{\rad}{\rho_{\text{rad}}}
\newcommand{\GW}{\rho_{\text{GW}}}
\newcommand{\M}{\rho_{\text{M}}}
\newcommand{\twoM}{\rho_{\text{2M}}}

\newcommand{\dotrad}{\dot{\rho}_{\text{rad}}}
\newcommand{\dotGW}{\dot{\rho}_{\text{GW}}}
\newcommand{\dotM}{\dot{\rho}_{\text{M}}}
\newcommand{\dottwoM}{\dot{\rho}_{\text{2M}}}

\title{\boldmath GUT-Scale Primordial Black Holes: Mergers and Gravitational Waves}

\author[a]{J. Luna Zagorac,}
\author[b]{Richard Easther}
\author[a]{and Nikhil Padmanabhan}
 
\affiliation[a]{Department of Physics,\\ Yale University, \\ New Haven, CT 06520, USA}
\affiliation[b]{Department of Physics,\\ University of Auckland, \\Private Bag 92019,\\ Auckland, New Zealand}

\emailAdd{luna.zagorac@yale.edu}
\emailAdd{r.easther@auckland.ac.nz}
\emailAdd{nikhil.padmanabhan@yale.edu}

\abstract{Tight constraints on the abundance of primordial black holes can be deduced across a vast range of masses, with the exception of those light enough to fully evaporate before nucleosynthesis. This hypothetical population is almost entirely unconstrained, to the point where the early Universe could pass through a matter-dominated phase with primordial black holes as the primary component. The only obvious relic of this phase would be Hawking radiated gravitons which would constitute a stochastic gravitational wave background in the present-day Universe, albeit at frequencies far beyond the scope of any planned detector technology. This paper explores the effects of classical mergers in such a matter dominated phase. For certain ranges of parameters, a significant fraction of the black holes merge, providing an additional, classical source of primordial gravitational waves. The resulting stochastic background typically has a lower amplitude than the Hawking background and lies at less extreme frequencies, but is unlikely to be easily detectable, with  a maximal present day density of $\Omega_{GW} \sim 10^{-12}$ and  frequencies between $10^{15}-10^{19} \, \rm{Hz}$. We also assess the impact of radiation accretion on the lifetimes of such primordial black holes and find that it increases the black hole mass by $\sim 14\%$ and the lifetimes by about $50\%$. However, this does not qualitatively change any of our conclusions.
}

\begin{document}
\maketitle
\flushbottom
%\linenumbers
\section{Introduction}
\label{sec:intro}

This paper examines the dynamics of a possible transient matter dominated phase in the very early Universe in which the Universe is dominated by primordial black holes \citep{hawking1971,Carr:1974nx,Anantua:2008am}, analysing the binary formation and coalescence and consequent gravitational wave production during this era. 
Primordial black holes (PBH) can form immediately after the Big Bang through the gravitational collapse of large primordial density fluctuations in the early Universe. We infer from observations that the early Universe was smooth on comoving scales corresponding to present-day astrophysical distances, but this does not constrain the portion of the primordial power spectrum relevant to PBH formation. Given that the measured portion of the primordial power spectrum is featureless and mildly red \cite{2018Planck}, naively extrapolating astrophysical results to smaller scales suggests that no PBH will form via direct collapse in the early Universe.\footnote{Mechanisms such as parametric resonance following inflation may dynamically magnify primary perturbations, providing alternative routes to PBH formation in scenarios for which primordial perturbations at all scales are small as they leave the horizon \cite{bassett2001inflationary}. In what follows we focus on PBH formed via the direct collapse of a near horizon-sized volume.} From the opposite perspective, however, limits on PBH abundance   constrain the primordial power spectrum at small length scales and  the extent to which the  scale-invariance of the power spectrum can be broken. These limits have implications for any physical mechanism responsible for generating the primordial perturbations, such as inflation \cite{green1997constraints,leach2000black,kohri2008black,peiris2008primordial}.

Black holes have finite lifetimes due to the emission of Hawking radiation. The Hawking temperature is inversely proportional to the mass and small black holes lose mass more rapidly that large ones. PBH can, in principle, range in size from the Planck mass to hundreds of solar masses and beyond; a PBH with an initial mass exceeding $10^{15}$ g survives until the present and is thus potentially detectable today.\footnote{The possibility that at least some of the LIGO detections reflect mergers of primordial rather than stellar black holes has attracted considerable attention \cite{Bird:2016dcv,2017JPhCS.840a2032G}.} Decaying black holes may disrupt nucleosynthesis, recombination, and reionization \cite{1992PhRvD..46..645B,mack2008primordial}, from which limits on populations of lighter PBH can be derived. However, a PBH population that decays well before nucleosynthesis has little impact on the canonical thermal history of the hot Big Bang and is thus largely unconstrained by observations \cite{2010PhRvD..81j4019C}. 

The absence of limits on populations of very short-lived PBH creates a range of phenomenological possibilities for the very early Universe, as discussed in Ref.~\citep{Anantua:2008am}. These include a transient matter dominated phase in which PBH are the dominant constituent of the Universe, as well as a stochastic gravitational wave background associated with Hawking radiated gravitons. This matter dominated stage would be part of the ``primordial dark age'' \cite{Boyle:2005se}, and could have an observable impact on the matching between primordial and present-day perturbation amplitudes \cite{Malquarti:2003ia,Easther:2006tv,Easther:2011yq} but the dynamics and consequences of this matter dominated phase are largely unexamined. In particular, the possible\textbf{\textbf{}} interactions between PBH in this epoch are entirely unexplored. 

In this paper we focus on black hole mergers and their consequences during a PBH dominated phase, in the primordial Universe to sharpen our understanding of the possible implications of such a scenario. Firstly, merged PBH can live up to eight times longer than their progenitors (since lifetime goes as the cube of the mass), creating the possibility that mergers extend the length of the matter dominated phase. Furthermore, mergers provide a second source of gravitational waves. Individual mergers will typically convert a larger fraction of progenitor restmass, about 5\%, into gravitational waves than Hawking decay, which only converts 2\% of evaporated mass into gravitons. \cite{Page:1976df} 

This paper is arranged as follows. In Section~\ref{sec:ultralightPBH} we survey the parameters governing the lifetime of PBH. In Section~\ref{sec:mergers} we present an estimate of the population and merger statistics of binary black holes in a Universe where PBH are uniformly and randomly distributed at formation, computing the binary capture rate induced by gravitational radiation during near-hyperbolic close encounters. These binaries can  undergo mergers and the resulting gravitational wave spectrum is computed in Section~\ref{sec:GWSpectrum}; we discuss our results in Section~\ref{Discussion}. We work in natural units where $\hbar = c = k_B = 1$ and the gravitational constant $G = M_P^{-2}$, where $M_P$ is the unreduced Planck mass.PBHPBH

\section{Ultralight PBH in the Early Universe} \label{sec:ultralightPBH}
%*******************************************************************
%
%
Primordial black hole formation in the early Universe is a well-studied topic
\cite{hawking1971,Carr:1974nx,1994caty.conf..132L, ivanov1998nonlinear,
 2000astro.ph..3027C,2007arXiv0708.3875H, carr2017primordial,
 carr2018primordial}. In what follows we make a number of simplifying assumptions to expose the underlying physics of this scenario, as we are interested in obtaining an estimate of the overall gravitational wave background, rather than its detailed form. Following the analysis of Ref.~\cite{Anantua:2008am}, we begin by assuming that the ultralight PBH have a mass equal to the
energy inside a Hubble volume at the moment of collapse, and that these black
holes form at the same time. This implies the black holes all have an initial mass of
\begin{equation}
M_i = \frac{4}{3} \pi \left(\frac{1}{H}\right)^3 \rho = \sqrt{\frac{3}{32\pi}}\frac{M_{P}^3}{E_{i}^2}, \label{initialmass}
\end{equation} 
where $\rho = E_i^4$ is the density for a radiation dominated Universe and $H^2 = 8\pi \rho/3 M_P^2$.
We also assume that the black holes are formed at rest and randomly distributed with an initial mass fraction
of $\Omega_{PBH, i} = \beta$. Our model is thus completely specified by two parameters, $E_i$ and
$\beta$. 
These black holes have a temperature 
\begin{equation} \label{BHtemp}
T= \frac{M_{P}^2}{8\pi M} \, 
\end{equation} 
and their mass decreases via Hawking radiation at a rate
\begin{equation} \label{Hawkmass}
\frac{dM}{dt} \Bigg|_H = - \frac{g}{30720\pi} \frac{M^4_{P}}{M^2} \, .
\end{equation}
The number of degrees of freedom into which the black hole
can radiate is denoted by $g$. The
resulting lifetime is then
\begin{equation} \label{Hawk_lifetime}
t_{H} = \frac{10240\pi}{g} \frac{M^3}{M^4_{P}} = \frac{240}{g} \sqrt{\frac{3}{2\pi}} \frac{M_{P}^5}{E_{i}^6}.
\end{equation}
Since the temperature of the black hole increases as its mass decreases, $g$ is a
dynamical quantity and should generically increase with time, but we assume it
is constant for simplicity. This is a good approximation since the mass, and
therefore the temperature of the black hole does not significantly change over
most of its lifetime. Because Hawking radiation is not strictly thermal and ``grey body'' corrections are more dramatic for higher-spin particles~\cite{Page:1976df}, $g$ is an effective number of degrees of
freedom. In general, the decay rate depends on the mix of spin-statistics in the
particle spectrum, but we absorb this ambiguity into the definition of $g$. For the PBH discussed here the temperature is always at the TeV scale or above, so Standard Model species may form a small subset of all relevant particle states. Furthermore, apart from the gravitational wave background described in \cite{Anantua:2008am}, we assume that all other radiated species equilibrate, leaving the early Universe in a thermal state after the PBH evaporate. In our numerical examples, we set $g = 1000$ for definiteness, but none of our conclusions depend strongly on this choice.

\begin{figure}[p]
 \centering
  \includegraphics[width=0.85\textwidth]{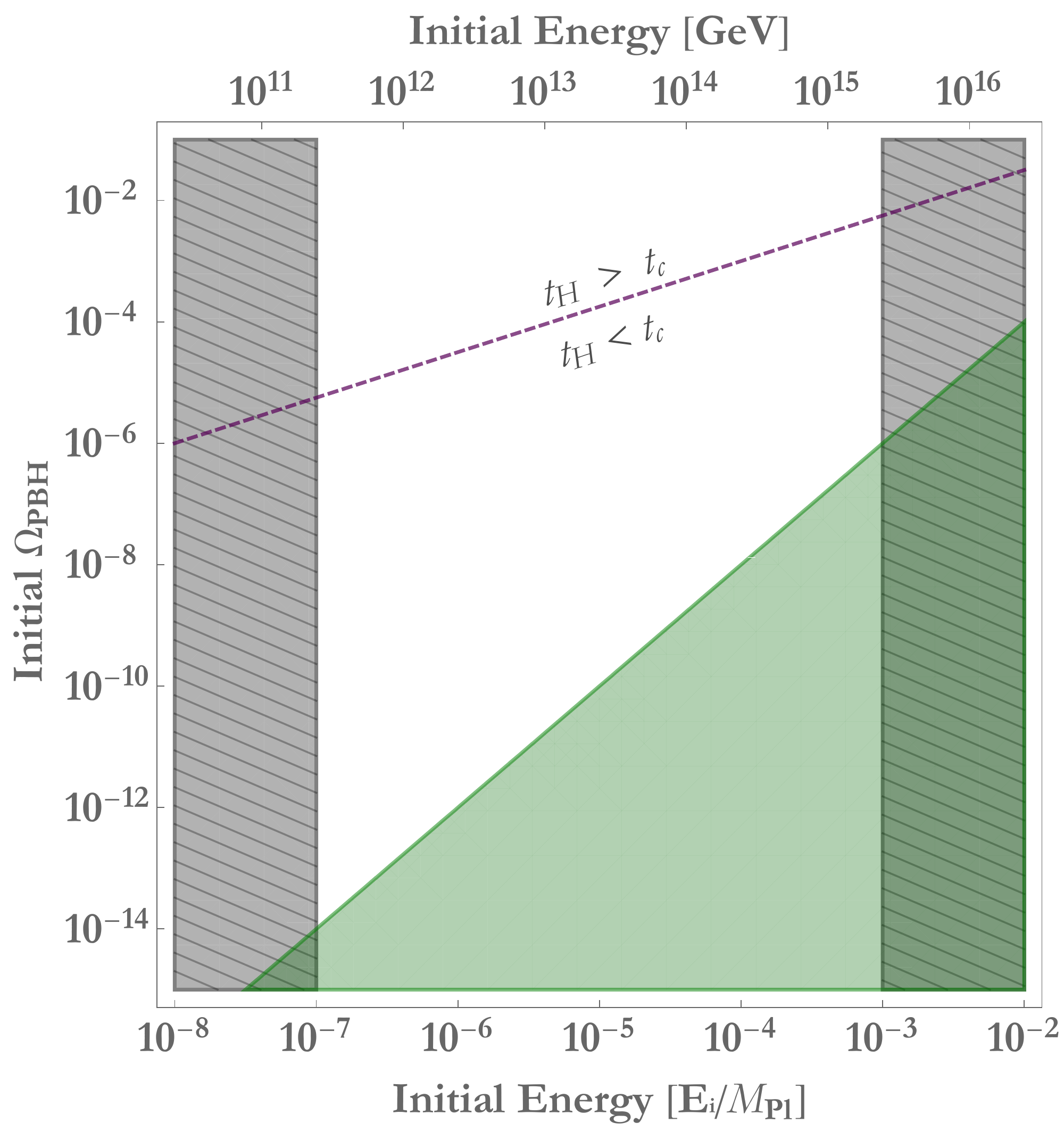} 
 \caption{ The $\beta - E_{i}$ parameter space. The gray dashed area on the right ($E_i > 10^{-3} M_P$) is excluded by CMB constraints if we assume a prior inflationary phase; the left hand area is excluded by requiring that PBH do not survive into BBN ($E_i < 10^{-7} M_P$). The green portion never undergoes a matter dominated phase. For regions above the dashed line the typical black hole separation yields a nominal merger timescale less than $t_H$; in the lower region only those black holes that form in relatively close proximity undergo mergers. } 
\label{fig:exclusion}
\end{figure}

Since radiation density scales as $a^{-4}$ while the matter scales like $a^{-3}$, the Universe can pass through an effective matter-dominated phase, with matter-radiation equality occurring when $a\sim 1/\beta$. Comparing the time to PBH-radiation equality with the PBH lifetime, the condition for the existence of a matter-dominated phase is \cite{Anantua:2008am}
\begin{equation} \label{beta-E}
\beta \geq \frac{1}{8} \sqrt{\frac{g}{15}} \frac{E^2_{i}}{M_P^2} .
\end{equation}
Matter-domination ends when the black holes fully evaporate, returning the Universe to radiation domination. Requiring this to occur before nucleosynthesis gives the constraint $t_H \leq 100  \text{ s}$. Taking $E_{i} \ge 10^{12}$~GeV results in $t_H = 29/g$ seconds, and an initial Hawking temperature of $T = 18.8$ TeV. This is actually an effective lower bound on $E_{i}$; the lifetime depends on $E_i^6$ so even a small softening of this limit allows PBH to survive into nucleosynthesis. \cite{2000PhRvD..61b3501K} Finally, if we assume an initial inflationary phase, bounds on the primordial gravitational wave background derived from microwave background data imply that $E_i \lesssim 10^{16}$ GeV, which is roughly $10^{-3} M_{P}$~\cite{2018Planck}, so we take this as a (tentative) upper bound. The resulting parameter space is summarized in figure \ref{fig:exclusion}.

Immediately after formation, the black holes are colder than the surrounding radiation bath, which has a temperature 
\begin{equation}
T_{\rm{univ}} = \Bigg( \frac{30 \rho_{rad}}{\pi^2 g}\Bigg)^{\frac{1}{4}} \, 
\end{equation}
so the accretion of radiation across the horizon can initially dominate the Hawking emission.\footnote{Note that the $g$ is in this equation is not fully consistent with the effective number of degrees of freedom in Hawking radiation expression, which includes the grey body terms; this difference can be ignored at the level of precision needed here.} While $T_{\rm{univ}} > T$, the mass of the black hole is increasing at the rate of 
\begin{equation} \label{accretion}
\frac{dM}{dt} \Bigg|_{acc} = f \times A \times \rho_{rad} = 16 \pi f \frac{M^2}{M_P^4} \rho_{rad} ,
\end{equation}
where $f$ is the accretion efficiency, and $A$ is the area of the black hole \cite{Nayak:2009wk}. We assume that when $T_{\rm{univ}} = T$ the Hawking radiation and accretion rates are roughly equal, $\dot{M}_{H} = \dot{M}_{acc}$ which implies that $f = 1/(4\pi)$, and in general $\dot{M} = \dot{M}_{H} + \dot{M}_{acc}$. Since the early Universe cools quickly, the overall mass-gain is not huge -- the maximal increase within the parameter range we consider is 14\%, but we include this term -- which was not accounted for in Ref. \cite{Anantua:2008am} -- for completeness.

%---------------------------------------------------------
%---------------------------------------------------------
%
%
\afterpage{\clearpage}
%
%
%%%%%%%%%%%%%%%%%%%%%%%%%%%%%%%%%%%%%%%%%%%%%%%%%%%%%%%%
\section{Mergers in the Early Universe}\label{sec:mergers}
%*******************************************************
Analyses of PBH mergers have recently enjoyed a renaissance thanks to suggestions that observed gravitational wave signals might be drawn from such a population rather than stellar remnants \cite{Bird:2016dcv, sasaki2016primordial, inomata2017inflationary, inayoshi2016gravitational, ali2017merger, chen2018merger}. We can map this directly to our problem\footnote{We ignore any clustering of PBH; the overall validity of this approximation is evaluated in Section \ref{Discussion}.} and we follow the simplified formalism developed in Ref. \cite{nakamura1997gravitational}. 
 We assume PBH are formed at rest and randomly distributed: in the absence of significant primordial non-Gaussianity, PBH will coalesce at random, uncorrelated locations. The first mergers  occur between pairs that happen to form in close proximity; the next-nearest neighbour is likely to be at a substantially greater distance and supplies a small perturbation preventing a head-on collision. However, gravitational radiation can lead to mutual capture during what would otherwise be a hyperbolic encounter in a Newtonian context, leaving the two PBH in a highly elliptical orbit. The larger accelerations experienced at periastron in an elliptical orbit enhance energy loss due to gravitational radiation relative to a circular binary with the same semimajor axis, shortening the coalescence time. 

 If PBH are randomly and uniformly distributed the expected number of PBH in any region depends only on its volume. When the condition in equation~\ref{beta-E} is satisfied there will be a transition between radiation and matter dominated growth in the primordial Universe, at a time $t = t_{eq}$. At this moment  $\Omega_{\mathrm{rad}} = \Omega_{PBH} = 1/2$, the total density is $\rho_{eq}$, and the characteristic comoving distance between PBH with mass $M$ is thus approximately 
\begin{equation} \label{x_bar}
\bar{x}_{eq} = \Bigg( \frac{2 M}{\rho_{eq}} \Bigg)^{\frac{1}{3}} .
\end{equation}
Given their creation mechanism, the PBH are initially at rest with respect to the background; they start falling toward each other when their contribution to the density of a sphere with diameter equal to their separation $x$ is greater than the average density of the Universe, or 
\begin{equation}
\rho_{PBH} = \rho_{eq} \Bigg( \frac{\bar{x}}{x a(t)} \Bigg)^3
\end{equation}
\begin{equation} \label{eq:avam}
a > a_m \equiv \Bigg( \frac{x}{\bar{x}} \Bigg)^3 \, .
\end{equation}
Here $a_m$ is the ($x$-dependent) value of the scale factor at which the ``density'' of PBH in a sphere of diameter $x$ containing two black holes exceeds the radiation density, at which point they decouple from the overall expansion of the Universe, and begin to fall toward each other. 

Before considering the coalescence rate of PBH binaries in detail, let us consider the key time scales that appear in the problem. The coalescence time for circular binaries is 
\begin{equation}\label{t_circular}
t_{c} = \frac{5}{16} \Bigg( \frac{\pi}{6}\Bigg)^{\frac{5}{6}} \frac{M_{P}}{E_{i}^2} \frac{1}{\beta^{\frac{16}{3}}} ,
\end{equation}
if the initial diameter is $\bar{x}$ at $t_{eq}$ \cite{peters1963gravitational,peters1964gravitational}. Comparing this to the Hawking lifetime (\ref{Hawk_lifetime}), 
\begin{equation} \label{mergerfraction}
\frac{t_{c}}{t_H} = \frac{g}{2304} \frac{\pi^{\frac{4}{3}}}{ 6^{\frac{1}{3}} } \frac{E_{i}^4}{M_{P}^4} \beta^{-\frac{16}{3}} \approx 1.1 \times 10^{28} \Bigg(\frac{E_i}{10^{-5} M_P} \Bigg)^4 \Bigg(\frac{10^{-9}}{\beta} \Bigg)^\frac{16}{3},
\end{equation}
divides our parameter space into two regions. For parameter choices with $t_H < t_c$ only those binaries that form in close proximity will merge before the black holes evaporate; conversely if $t_H > t_c$ there is the potential for a nontrivial number of second (and higher) generation mergers.  These regions are illustrated in figure \ref{fig:exclusion}. 

In addition, we can compare the circular coalesence timescale $t_c$ with the Hubble time ($1/H$) at matter-radiation equality 
\begin{equation}
t_{c} H_{eq} = \frac{5}{24} \Bigg( \frac{\pi^4}{6} \Bigg)^{\frac{1}{3}} \frac{1}{\beta^{10/3}}
\end{equation}
showing that the merger time will always be long compared to the initial Hubble time unless $\beta$ is very close to unity. This means that matter dominated phase will be roughly $1/\beta^{10/3}$ times longer than the Hubble time at equality, from which we deduce that the matter dominated Universe grows by a factor of $ t^{2/3} = 1/\beta^{20/9}$. We can then imagine a second generation of mergers beginning at an effective separation of $\bar{x}/\beta^{20/9}$, with their initial number reduced by a factor of 2. This extends the naive timescale for all mergers to complete by $2^{4/3}/\beta^{80/9}$. Even for a high value of $\beta = 0.1$, this increase is on the order of $10^9$, which swamps the roughly eightfold increase in lifetime gained by second generation black holes. Consequently, it seems that even if $t_H > t_c$, a ``cascade'' of mergers is unlikely. However, this argument assumes an initially uniform, random distribution of PBH and the second generation mergers could be facilitated by clustering and N-body interactions, a full treatment of which is beyond the scope of this analysis. However, the authors of Ref. \cite{2019JCAP...02..018R} explored the N-body merger dynamics of younger, more massive black holes, and found that the merger rate is suppressed through the disruption of early-formed binaries by nearest neighbors, making subsequent rounds of mergers even more unlikely.   

To estimate the merger rate in more detail we follow the approach of Ref. \cite{nakamura1997gravitational}. 
Consider a pair of black holes with comoving separation $x$, $0 < x < \bar{x}$, which breaks free from the Hubble expansion when $a=a_m$. Their initial separation sets the major axis of their orbit, $\alpha_m$: %
\begin{equation} \label{alpha}
\alpha_m = xa_m = \frac{x^4}{ \bar{x}^3} \, .
\end{equation} 
We estimate the minor axis $\beta_{m}$ as the (tidal acceleration) $\times$ (free fall time)$^2$, giving 
\begin{equation}
\beta_{m} = \frac{G M xa_m}{(ya_m)^3} \frac{(xa_m)^3}{GM} = \Bigg( \frac{x}{y}\Bigg)^3 \alpha_m, \label{at2}
\end{equation}
where $y$ is the separation of the next-to-nearest neighbour to the center of mass of the binary.\footnote{We follow the presentation of Ref. \cite{nakamura1997gravitational}. One could add a factor of $\frac{1}{2}$ to equation \ref{alpha}, since this is technically the major (not semimajor) axis and this would induce a factor of $\frac{1}{2}$ in equation \ref{at2} (i.e. so that reads $\frac{1}{2}at^2$ instead of $at^2$). These factors, along with a proper normalization of \ref{probdistribution} leave the final probability distribution unchanged. } From here, it is straightforward to work out the ellipticity,
\begin{equation}
e = \sqrt{1 - (x/y)^6}.
\end{equation}
We must now decide on the form of the distribution of $x$ and $y$. \cite{nakamura1997gravitational} assume that $x$ and $y$ are uniformly distributed in 3D, resulting in
\begin{eqnarray}
\begin{aligned} \label{probdistribution}
f(\alpha_m,e)d\alpha_m de &= 18 \frac{x^2 y^2}{ \bar{x}^{6}}dxdy \\
&= \frac{3}{2}\frac{\alpha_m^{\frac{1}{2}}e}{\bar{x}^{\frac{3}{2}}(1-e^2)^{\frac{3}{2}}}d\alpha_m de \label{ellipse_prob} .
\end{aligned}
\end{eqnarray}
While this assumption cannot be strictly correct, \cite{1998PhRvD..58f3003I} show that it is
accurate to within a factor of 2. Given our other assumptions, we adopt the simpler formulation in \cite{nakamura1997gravitational}. As we'll see below, using the more
accurate formulation will not change our conclusions.

Elliptical binaries coalesce in a time \cite{peters1963gravitational,peters1964gravitational}
\begin{equation}
t = t_c \Bigg( \frac{\alpha_m}{\bar{x}}\Bigg)^4 (1 - e^2)^{\frac{7}{2}}
\end{equation}
where $t_c$ is the circular coalescence time, equation \ref{t_circular}. Substituting into equation \ref{ellipse_prob} and integrating yields the distribution of binary lifetimes:
\begin{equation} \label{f(t)dt}
f(t)dt = \frac{3}{29} \Bigg[ \Bigg( \frac{t}{t_c}\Bigg)^{\frac{3}{37}} - \Bigg( \frac{t}{t_c}\Bigg)^{\frac{3}{8}} \Bigg] \frac{dt}{t}.
\end{equation}
This suggests that all black holes have merged when $t=t_c$; however this analysis is only strictly applicable in the tail of the distribution for pairs with very small initial separations, given that it is based upon the lifetime of a tightly bound binary that is far from its nearest neighbours. However, as noted above, $t_c$ sets the timescale over which we would expect a significant fraction of the population to undergo mergers. Note that we have not accounted for nontrivial three-body interactions, which could leave two participants tightly bound while the third is accelerated to a large (relative) velocity. A full treatment of this such interactions would require a full $N$-body simulation. 

%%%%%%%%%%%%%%%%%%%%%%%%%%%%%%%%%%%%%%%%%%%%%%%%%%%%%%%%%%%%%%%%%%
\section{Gravitational Wave Spectrum} \label{sec:GWSpectrum}

To calculate the density of gravitational waves, we set up the following system of equations. The two parameters which scale as matter are denoted $\M$ and $\twoM$ representing the density of ``first generation'' black holes with original mass $M,$ and $\twoM$ accounts for the post-merger population, respectively. The two parameters which scale as radiation are denoted as $\rad$, representing the density of the Universe in radiation, and $\GW$, which represents the gravitational radiation produced by mergers. Accounting for the expansion of the Universe and the source/sink terms, and recalling that $\M = n(t)M(t)$, we have the network of equations

\begin{align}
\dotM & = \dot{n}(t)M(t) + n(t)\dot{M}(t) \nonumber \\ 
 & = -3 \frac{\dot{a}}{a} \M - \sigma(t)\M + \M \frac{\dot{M}}{M} \label{rhoBHdot}\\
\label{rhoMdot}
\dottwoM &= -3 \frac{\dot{a}}{a} \twoM + (1 - \epsilon) \sigma(t) \M \\ %+ \twoM \frac{\dot{M}}{M}\\
\label{rhoRADdot}
\dotrad &= -4 \frac{\dot{a}}{a} \rad - \M \frac{\dot{M}}{M} \\%- \twoM \frac{\dot{M}}{M}\\
\dotGW &= -4 \frac{\dot{a}}{a} \GW + \epsilon \sigma(t) \M \label{rhoGWdot} \\
\frac{\dot{a}}{a} &= \Bigg[ \frac{8\pi}{3M_P^2} ( \M + \twoM + \rad + \GW) \Bigg]^\frac{1}{2}  \label{adot} \\
\dot{M} &= - \frac{g}{30720\pi} \frac{M^4_{P}}{M^2} + 4 \frac{M^2}{M_P^4} \rho_{rad} . \label{mass_differential}
\end{align}

%---------------------------------------------------------
\begin{figure}[p]
\centering
\includegraphics[width=1.0\textwidth]{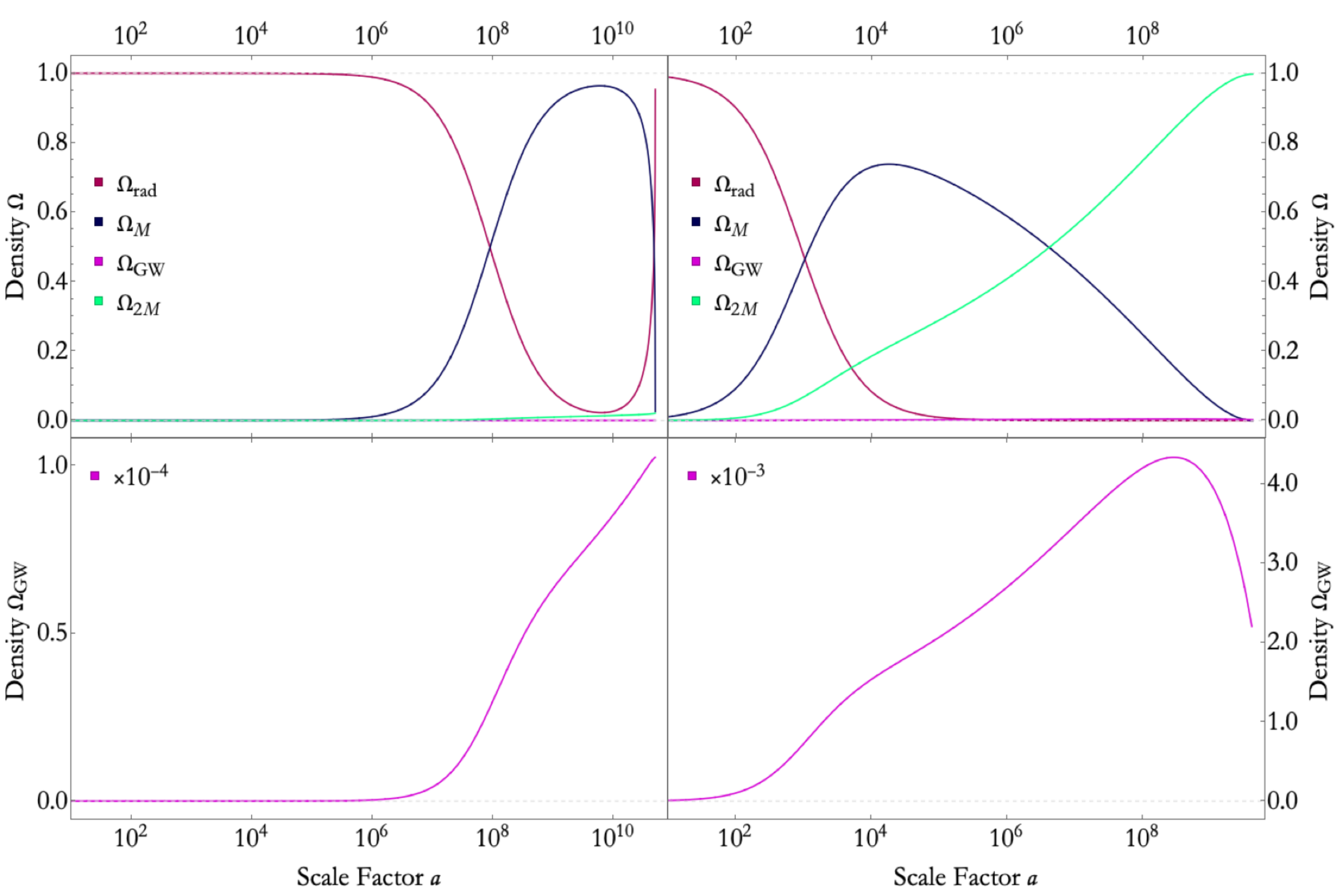} \\
\caption{The evolution of $\Omega_{rad}$, $\Omega_{M}$, $\Omega_{2M}$, and the merger-generated gravitational wave background, $\Omega_{GW}$. On the upper left, we plot the behavior of the system for $E_i = 10^{-5} M_{P}$ and $\beta = 10^{-8}$, for which $t_H < t_c$. On the right, we plot the behavior for $E_i = 10^{-5} M_{P}$ and $\beta = 10^{-3}$, for which $t_c < t_H$, and so most binaries coalesce. For this case, we stop the integration when
the first generation of mergers has completed. The turnover in the density
of gravitational waves is due to their redshifting during the matter
dominated era. } \label{fig:Omegas_vs_loga}
\end{figure} 
\afterpage{\clearpage}
%---------------------------------------------------------
%

%
The last term in equation \ref{rhoBHdot} accounts for both Hawking radiation and accretion; the sign convention ensures that $\dot{M}(t)$ is positive during accretion and negative during evaporation (see equation \ref{mass_differential}). The first two terms in equation \ref{rhoBHdot} arise from the changing number density of PBH: the first indicates the density of black holes scales as matter in the expanding Universe, and the second accounts for black hole mergers, where
\begin{equation}
\sigma(t) = f(t) \Big[ 1 - \int_0^t f(t) \Big]^{-1}
\end{equation}
 is the normalized fractional change in number density per unit time\footnote{The lower bound of the integral of $f(t)$ here should technically be PBH formation time $t_0$; compared to any other timescale in the problem, this is essentially zero.} derived from the merger time probability distribution of equation~\ref{f(t)dt}. We assume that a fixed fraction $\epsilon$ of the rest mass is converted into gravitational waves by the merger while the remaining $1 - \epsilon$ comprises the newly formed black hole; in what follows we take $\epsilon = 0.05$ \cite{2001astro.ph..8028P, 2016PhRvL.116m1102A}.  The first terms in equations \ref{rhoRADdot} and \ref{rhoGWdot} accounts for the usual dilution of radiation in an expanding Universe. Finally, the evolution of the scale factor is given by equation \ref{adot}.

The initial conditions are $M(t_0) = M_i$, $\M(t_0) = \beta E_i^4$, $\rad(t_0) = (1-\beta)E_i^4$, $a(t_0) = 1$, and $\rho_{2M}(t_0) = \GW(t_0) = 0$. We evolve the equations until $M(t)$ becomes $10^{-5} M_i$, with $M_i$ given by equation \ref{initialmass}; we cannot use the analytic expression for $t_H$ (equation \ref{Hawk_lifetime}) as a stopping condition since it does not account for radiation accretion. Illustrative examples for two sets of parameters are shown in figure \ref{fig:Omegas_vs_loga}.

We have made a number of approximations here. Firstly, we are ignoring Hawking radiation from the more massive black holes; the practical advantage of this is that because we are assuming the initial PBH population has identical masses, the Hawking emission per unit mass is only a function of time. However, the masses of the black holes produced during mergers will depend on when the mergers take place, which would complicate the computation of the emission rate. As we are only looking at classically produced gravitational waves and since secondary mergers are likely to be rare, this approximation is not unreasonable. We also ignore the impact of decreasing black hole masses on merger timescales; since the mass initially changes slowly, this is a reasonable approximation and will lead to an overestimate of gravitational wave production during the last few Hubble times. However, since the Universe grows by a very large factor during any PBH dominated phase this simplification is also relatively benign.

\afterpage{\clearpage}
%%%%%%%%%% 

\begin{figure}[!ht]
\centering
\includegraphics[width=0.95\textwidth]{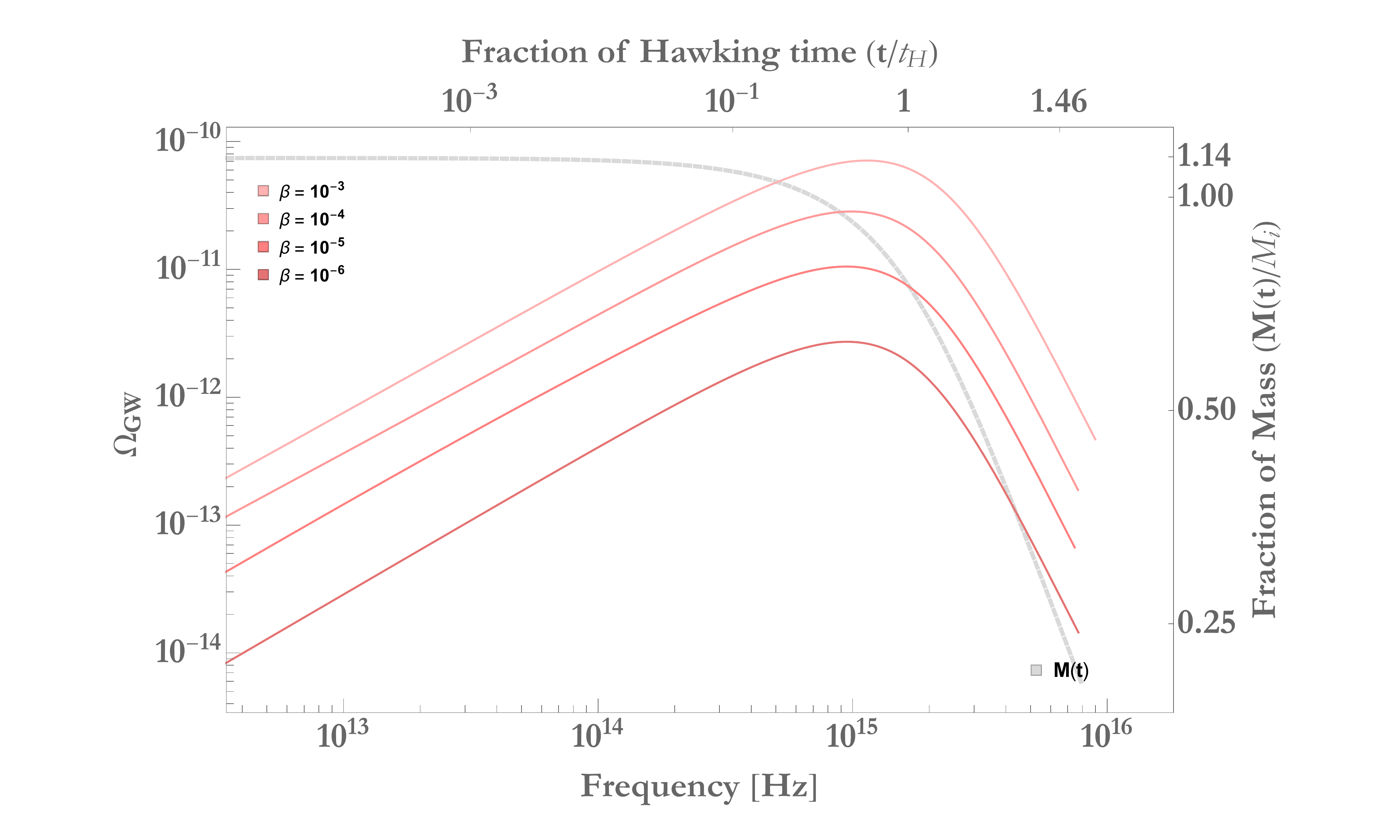}\label{fig:Ei3}
\hfill
\includegraphics[width=0.95\textwidth]{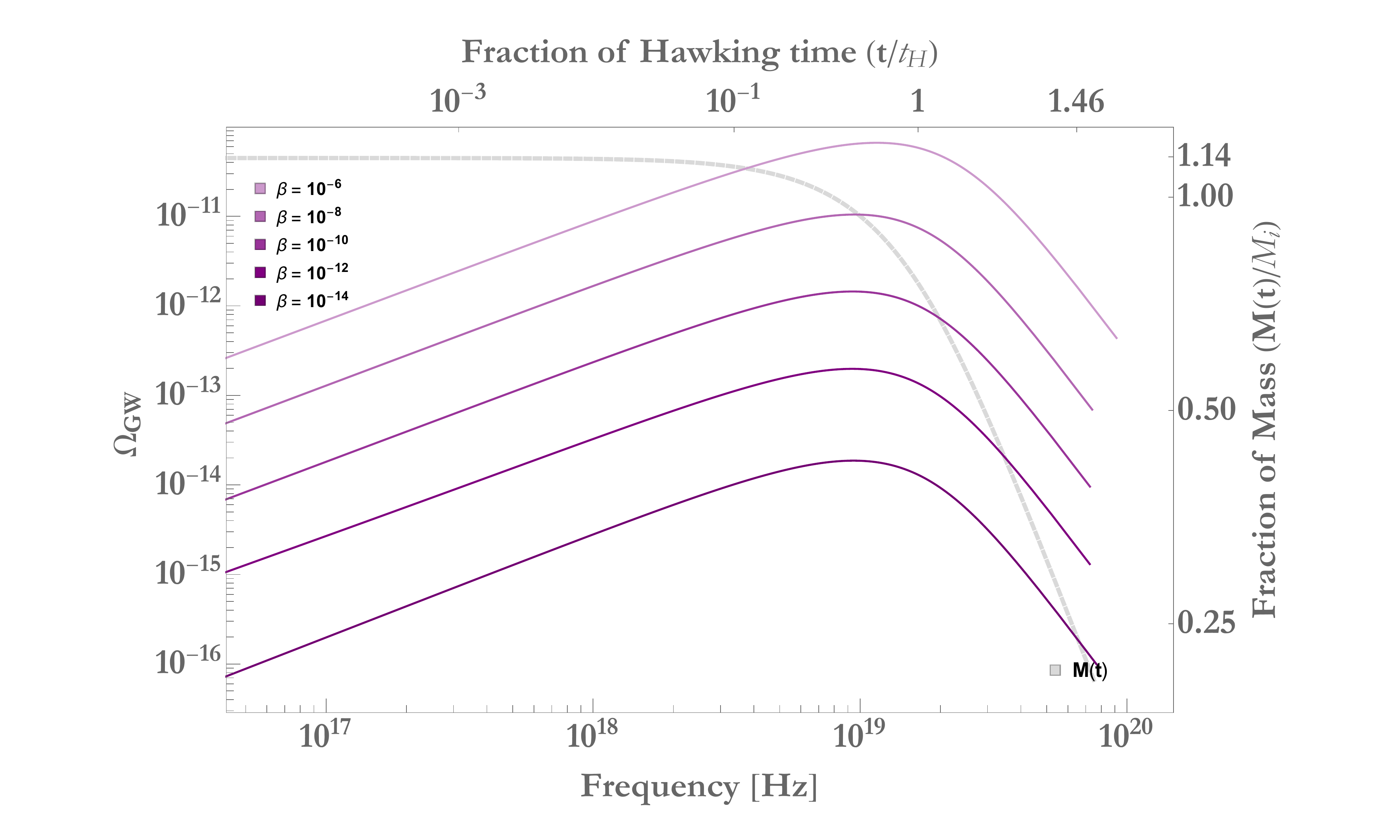}\label{fig:Ei7}
\caption{Present day spectra of gravitational waves from coalescing black holes for parameter choices where most binaries do not coalesce before evaporating, $t_H < t_c$. The top panel has $E_i = 10^{-3} M_P$, while the lower panel has $E_i = 10^{-7} M_P$, and the values of $\beta$ are shown in the legends. The fractional mass $M(t)/M_i$ of the initial black holes at the moment the corresponding gravitational waves are produced is shown in gray and labeled on the right axis, and the fraction of the initial Hawking time (which including the impact of early coalescence) is plotted on the top axis. Adjusting $E_i$ changes the frequency range, while $\beta$ controls the amplitude of the spectrum. }
\label{fig:spectra_today}
\end{figure} 
 
%%%%%%%%% 

To compute the spectrum of classically produced gravitational waves, we assume mergers produce a monochromatic gravitational wave signal. This very strong approximation is workable given that the Universe typically expands by many orders of magnitude during the black hole dominated phase, ``smearing'' the spectrum of the resulting stochastic background. We build the overall present-day spectrum by redshifting the produced gravitational waves after they are produced. The peak frequency (at the time of merger) can be deduced from Ref.~\cite{2001astro.ph..8028P, 2016PhRvL.116m1102A} 
\begin{equation}
\nu_{\rm{rest}} = \frac{1}{4\pi^2 G M(t)}. 
\end{equation}
The frequency of the emitted GW thus scales as
\begin{equation}
\nu (t) = \frac{a(t)}{a(t_e)} \nu_{\rm{rest}}, \label{nu}
\end{equation}
where $t_e > t_H$ is the time of PBH evaporation when accounting for accretion.
Solving equations \ref{rhoBHdot} - \ref{mass_differential} yields an expression for the total $\Omega_{GW}(t)$; to obtain the gravitational radiation at a given frequency we follow Refs~\cite{easther2006stochastic, price2008stochastic}:
\begin{equation}
\Omega_{\text{GW}}(\nu) = \frac{1}{\rho_c(t_e)} \frac{d\GW}{d\ln(\nu)}.
\end{equation}
We transform this into a more convenient form
\begin{equation}
\Omega_{\text{GW}}(\nu) = \frac{\dotGW}{\rho_c(t_e)} \frac{\nu}{\dot{\nu}},
\end{equation}
where $\dot{\rho}_{GW} = \epsilon \sigma(t) \M(t)$ is the contribution to the gravitational wave background at time $t$, $\rho_c(t_e)$ is the critical density at PBH evaporation, and $\nu / \dot{\nu}$ can be calculated from equation \ref{nu}. To calculate the present-day $\Omega_{GW}$ from mergers we assume that the Universe is radiation dominated from the moment the primary black hole population decays through to the moment matter-radiation equality prior to recombination. We redshift this background to present day values $\Omega_r$ via the relationship \cite{easther2006stochastic}
\begin{equation}
\Omega_{\text{GW}}^{\text{today}} = \Bigg( \frac{g_{0}}{g_*} \Bigg)^{4/3} \frac{\Omega_{\rm{rad}}^{\text{today}}}{\Omega_{\rm{rad}}^{\rm{evap}}} \Omega_{\text{GW}}^{\rm{evap}}.
\end{equation}
We take $g_* = 1000$, $g_{0} = 3.91$ \cite{husdal2016effective}, $\Omega_{\rm{rad}}^{\rm{today}} = 8.24 \times 10^{-5}$, and $\Omega_{\rm{rad}}^{\rm{evap}} = 1$ and assume that the temperature of the Universe immediately after evaporation is 
\begin{equation}
T_{e} = \frac{30 \rad(t_e)}{(\pi^2 g_*)^{\frac{1}{4}}}
\end{equation}
which yields the redshifted frequencies
\begin{equation}
\nu^{\text{today}} = \Bigg( \frac{g_*}{g_{0}}\Bigg)^{\frac{1}{3}} \Bigg( \frac{T_{\text{CMB}}}{T_e}\Bigg) \nu.
\end{equation} 

Representative results for the region of parameter space where $t_H < t_c$ are shown in figure \ref{fig:spectra_today}. The rising part of the spectrum is generated when $t \le 0.5 t_H$; in this regime, after accounting for accretion, the PBH masses are relatively unchanged from their initial values. This phase can see substantial gravitational wave production, but the gravitational waves produced by earlier mergers undergo a greater redshift during the primordial matter dominated phase, reducing their amplitude in the present-day Universe. Conversely, the falling portion of the spectrum is generated when $t > 0.5 t_H$, and drops because the masses of coalescing black holes are now significantly reduced by evaporation. It is evident that $E_i$ controls the frequency range, while $\beta$ controls the amplitude of the spectrum.

%
%---------------------------------------------------------
\begin{figure}[!t]
 \centering
 \includegraphics[width=0.95\textwidth]{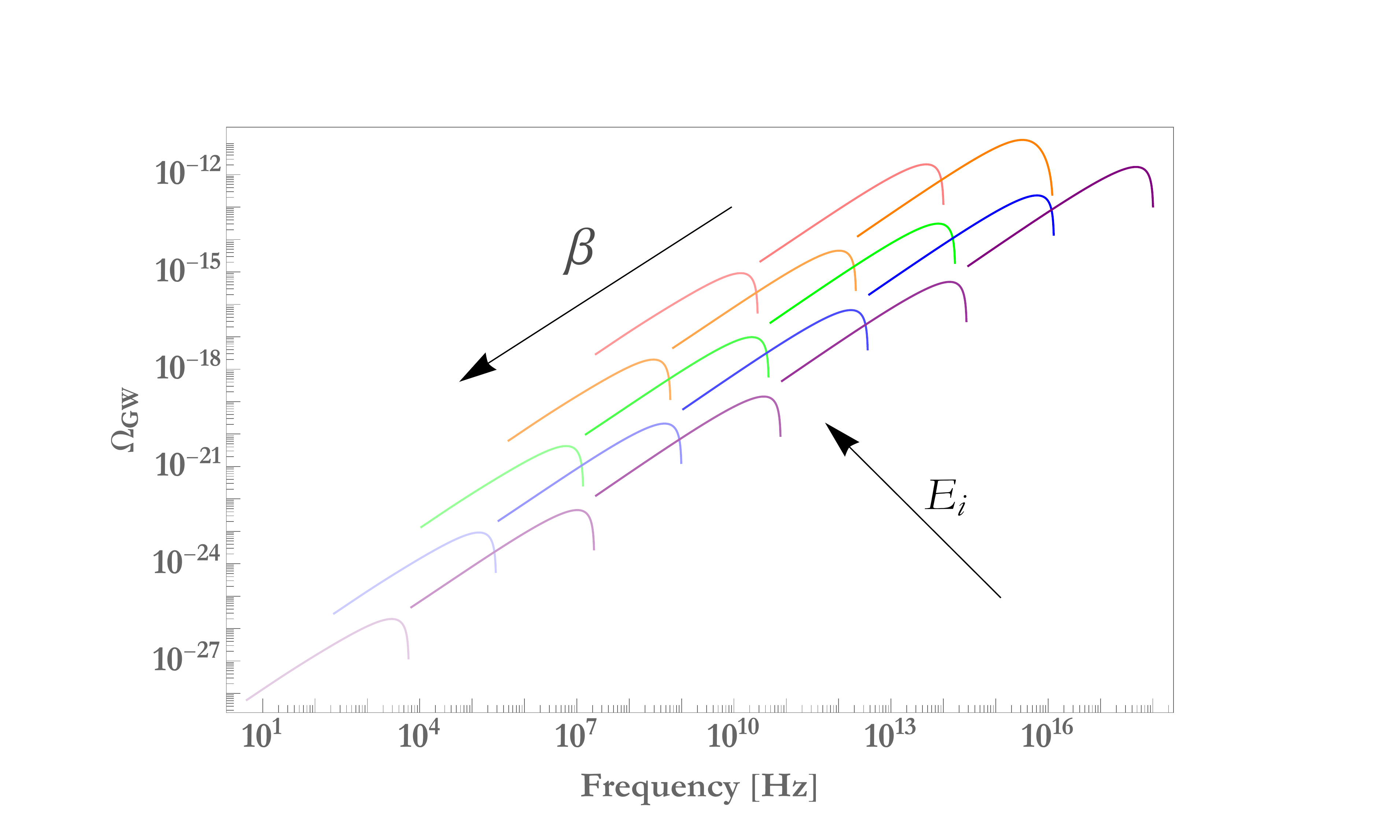} 
 \caption{Gravitational wave spectra from coalescing black holes for regions of parameter space when most binaries coalesce, $t_c<t_H$. Each color corresponds to a single value of $E_i$ as identified in figure \ref{fig:spectra_today}, with the addition of orange for $E_i = 10^{-4} M_P$, green for $E_i = 10^{-5} M_P$, and blue for $E_i = 10^{-6} M_P$. The highest $\beta$-values for each $E_i$ are $\beta = 10^{-1}$, and each darker hue moving to the right represents a factor of 10 decrease in $\beta$. In order to produce a spectrum with LIGO-range frequencies (10 Hz - 1 kHz), about 10\% of the Universe must start out in the form of black holes and the resulting amplitude is undetectably small.
 } 
\label{fig:spectratoday_RR}
\end{figure}
%---------------------------------------------------------

 %
For parameter choices for which $t_H > t_c$, we evolve our coupled differential equations until $t = t_c$, where the majority of mergers will have concluded. Then, we redshift from $a(t_c)$ to $a(8t_H)$, making the approximation that the Universe was matter dominated during this time. We then redshift from $t_c$ to today as previously, using 
\begin{equation}
\rho_{rad}(8t_H) = \Bigg ( \frac{a(8t_H)}{a(t_c)} \Bigg)^{-\frac{2}{3}} \rho_{2M}(t_c),
\end{equation}
since at $t_c$ all the density of the Universe (save the produced gravity waves) is contained in $\rho_{M}$, which Hawking evaporates into radiation at $8t_H$. The resulting spectra are shown in figure \ref{fig:spectratoday_RR}. The additional redshift from the longer matter dominated phase can significantly reduce the frequency, but at the cost of also lowering the density of gravitational waves. This makes it possible to produce spectra in the LIGO frequency band through mergers, but the corresponding $\Omega_{GW}$ is many orders of magnitude below even the most optimistic sensitivities, and requires $\beta$ to be on the order of 0.1, or greater.

Figure \ref{fig:spectratoday_RR} allows us to guess at the behavior when multiple generations of mergers take place. If
we imagine that the first generation of mergers completes rapidly, we can treat the resulting
population of black holes as being approximately coeval and simply iterate our calculation.
As Figure 4 shows, the radiation from the prior generations is heavily redshifted and will contribute
a very small fraction to the final radiation density. Ultimately, this process gets truncated by the
expansion of the Universe during the matter dominated era, which dilutes the density of PBH (as described
previously).

\begin{figure}[p]
 \centering
  \includegraphics[width=0.95 \textwidth]{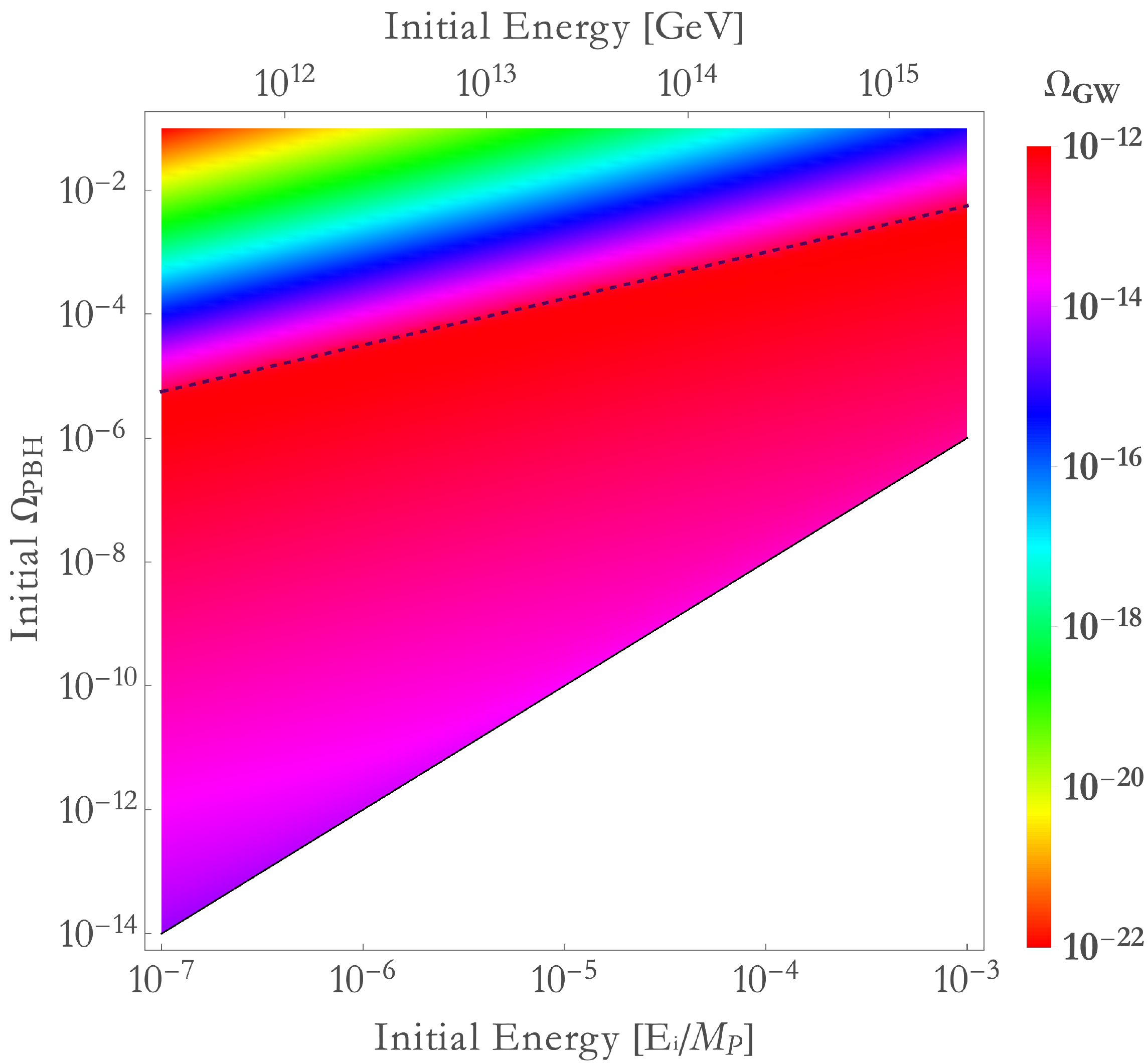} 
 \caption{ Maximum value of $\Omega_{GW}$ today, for parameter choices with which the Universe undergoes a matter dominated phase and  are consistent with other broad physical constraints.  After redshifting, the maximal amplitude is found for a parameter choice for which $t_c$ is approximately equal to $t_H$, around and below the dashed line. Although the upper region has a majority of black holes merging, it contributes less to $\Omega_{GW}$ because the additional redshifting from completion of merger to Hawking evaporation is significant.   } 
\label{densityplot} 
\end{figure}

Finally, figure \ref{densityplot} shows the present-day value of $\Omega_{GW}$ as a function of $\beta$ and $E_i$. For scenarios where $t_H> t_c$ we can estimate $\Omega_{GW}$ with reasonable confidence; but as noted above when $t_H < t_c$ we make a guesstimate based on an assumption that mergers continue until $t=t_c$ and ignore secondary mergers.  In these scenarios, our assumptions imply that a larger value of $\beta$ leads to a smaller present-day value of $\Omega_{GW}$; in these cases the total production of gravitational waves is similar, but when the mergers complete relatively early in the matter dominated phase, the subsequent dilution of the radiation component is greater. The integrated background of gravitational waves can, in principle, be detected as an additional radiation component. These are often parametrized as a fractional increase in the number of effective neutrino species, with $\Delta N_{\rm eff} = 1$ corresponding to $\Omega_{GW} h^2 = 5.6 \times 10^{-6}$ today \cite{2006PhRvL..97b1301S}. However, as is clear from this figure, even the most optimistic scenarios would not be detectable today.

\afterpage{\clearpage}

\section{Discussion}\label{Discussion}

We have analyzed primordial black hole (PBH) mergers and resulting gravitational wave production during a transient matter dominated phase in the primordial Universe, when the matter content of the Universe predominantly consists of PBH. We estimate the merger rate during this phase, and for some parameter values find that there will be numerous interactions between the PBH. These mergers generate gravitational waves, which will survive through to the present day. However, the characteristic frequency of these gravitational waves is typically huge -- well above $10^{10}$ Hz in the present epoch for most parameter choices. In fact, the only candidate gravitational wave background with a higher frequency are the Hawking radiated gravitons produced as these black holes decay, which was described in Ref.~\cite{Anantua:2008am}. However, the Hawking radiated background achieves higher values of $\Omega_{GW}$ than the merger-generated gravitational waves, but peaks at higher frequencies and falls off much more quickly than the merger background to the left of its peak. The background from mergers can thus dominate at lower frequencies, but the present-day density of gravitational waves is  very low in this regime.  

Our treatment makes a number of strong assumptions, so our results should be seen as an order of magnitude estimate rather than a detailed prediction, but this is adequate for present purposes. In particular, we assume that the initial PBH population has a monochromatic mass function (where $M = M_{Horizon}$) at time $t_0$ and that the initial spatial distribution of black holes is uniform and randomly distributed. Ref. \cite{2017JCAP...09..037R} maintains the assumption of uniformity while accounting for the possibility of a log-normal mass distribution for a younger population of PBH. Naively, this skews the average mass to a slightly higher number, corresponding to lower $E_i$ values and thus leading to somewhat lower $\Omega_{GW}$ values for our spectra. In addition, while we include the impact of radiation accretion on the black hole mass, we do not allow for the additional dissipation induced as rapidly moving black holes encounter a ``headwind'' of radiation. We also assume that the resulting gravitational waves are radiated monochromatically when computing the spectrum, but the continuous merger processes in combination with the large growth that occurs during the matter dominated softens the impact of this assumption. 
 
The major omission in our analysis is the impact of clustering, both at the point of formation, which will depend on the detailed physical mechanism associated with PBH formation \cite{chisholm2006clustering,chisholm2011clustering,2018arXiv180505912A,desjacques2018spatial,ballesteros2018merger}, and as a result of large scale clustering which grows during the matter dominated phase. In scenarios where only a small fraction of the total black hole population undergoes mergers, our treatment is likely to be sufficiently robust if the assumption of a uniform, random initial distribution is reasonable. However, in scenarios where the typical coalescence time is less than the Hawking time our formalism predicts that all black holes will undergo one merger but discounts the possibility of multiple mergers. It is worth emphasizing that our calculated backgrounds are a number of orders of magnitude away from the expected
detection limits. Furthermore, when merging is efficient, we observe that the gravitational radiation from early
mergers is redshifted away. This suggests that clustering is unlikely to qualitatively change this picture.

The key result of this paper is to demonstrate that any transient PBH dominated phase in the early Universe is likely to support mergers. In almost all cases, while these mergers further enrich the possible phenomenology of the ``primordial dark age'' \cite{Boyle:2005se,Malquarti:2003ia,Easther:2006tv,Easther:2011yq} they are unlikely to prolong this PBH dominated phase to the point that it would disrupt nucleosynthesis and thus cannot further constrain the PBH populations or the underlying mechanisms that gives rises them, with the caveat that the full N-body dynamics of this epoch are still to be explored. That said, these mergers would generate a stochastic background of gravitational waves that can presumably survive through to the present day. However, these gravitational waves typically have very high frequencies, and would be subdominant relative to the Hawking radiated gravitational wave background produced as the PBH decay. 

\acknowledgments

Preliminary results of this paper were presented at the 2018 Tri-Institute Summer School on Elementary Particles (TRISEP). LZ would like to acknowledge the Perimeter Institute and the Yale Conference Travel Fellowship for their financial support in aid of attending the summer school. The authors would like to thank Kazunori Kohri and Ville Vaskonen for useful discussions.


\bibliographystyle{JHEP}
\bibliography{bibliography.bib}

\providecommand{\href}[2]{#2}\begingroup\raggedright\begin{thebibliography}{10}

\bibitem{hawking1971}
S.~{Hawking}, \emph{{Gravitationally collapsed objects of very low mass}},
  \href{https://doi.org/10.1093/mnras/152.1.75}{\emph{\mnras} {\bfseries 152}
  (1971) 75}.

\bibitem{Carr:1974nx}
B.~J. {Carr} and S.~W. {Hawking}, \emph{{Black holes in the early Universe}},
  \href{https://doi.org/10.1093/mnras/168.2.399}{\emph{\mnras} {\bfseries 168}
  (1974) 399}.

\bibitem{Anantua:2008am}
R.~Anantua, R.~Easther and J.~T. Giblin, \emph{{GUT-Scale Primordial Black
  Holes: Consequences and Constraints}},
  \href{https://doi.org/10.1103/PhysRevLett.103.111303}{\emph{Phys. Rev. Lett.}
  {\bfseries 103} (2009) 111303}
  [\href{https://arxiv.org/abs/0812.0825}{{\ttfamily 0812.0825}}].

\bibitem{2018Planck}
{Planck Collaboration}, N.~{Aghanim}, Y.~{Akrami}, M.~{Ashdown}, J.~{Aumont},
  C.~{Baccigalupi} et~al., \emph{{Planck 2018 results. VI. Cosmological
  parameters}}, {\emph{arXiv e-prints} (2018) arXiv:1807.06209}
  [\href{https://arxiv.org/abs/1807.06209}{{\ttfamily 1807.06209}}].

\bibitem{bassett2001inflationary}
B.~A. {Bassett} and S.~{Tsujikawa}, \emph{{Inflationary preheating and
  primordial black holes}},
  \href{https://doi.org/10.1103/PhysRevD.63.123503}{\emph{\prd} {\bfseries 63}
  (2001) 123503} [\href{https://arxiv.org/abs/hep-ph/0008328}{{\ttfamily
  hep-ph/0008328}}].

\bibitem{green1997constraints}
A.~M. {Green} and A.~R. {Liddle}, \emph{{Constraints on the density
  perturbation spectrum from primordial black holes}},
  \href{https://doi.org/10.1103/PhysRevD.56.6166}{\emph{\prd} {\bfseries 56}
  (1997) 6166} [\href{https://arxiv.org/abs/astro-ph/9704251}{{\ttfamily
  astro-ph/9704251}}].

\bibitem{leach2000black}
S.~M. {Leach}, I.~J. {Grivell} and A.~R. {Liddle}, \emph{{Black hole
  constraints on the running-mass inflation model}},
  \href{https://doi.org/10.1103/PhysRevD.62.043516}{\emph{\prd} {\bfseries 62}
  (2000) 043516} [\href{https://arxiv.org/abs/astro-ph/0004296}{{\ttfamily
  astro-ph/0004296}}].

\bibitem{kohri2008black}
K.~{Kohri}, D.~H. {Lyth} and A.~{Melchiorri}, \emph{{Black hole formation and
  slow-roll inflation}},
  \href{https://doi.org/10.1088/1475-7516/2008/04/038}{\emph{Journal of
  Cosmology and Astro-Particle Physics} {\bfseries 2008} (2008) 038}
  [\href{https://arxiv.org/abs/0711.5006}{{\ttfamily 0711.5006}}].

\bibitem{peiris2008primordial}
H.~V. {Peiris} and R.~{Easther}, \emph{{Primordial black holes, eternal
  inflation, and the inflationary parameter space after WMAP5}},
  \href{https://doi.org/10.1088/1475-7516/2008/07/024}{\emph{Journal of
  Cosmology and Astro-Particle Physics} {\bfseries 2008} (2008) 024}
  [\href{https://arxiv.org/abs/0805.2154}{{\ttfamily 0805.2154}}].

\bibitem{Bird:2016dcv}
S.~Bird, I.~Cholis, J.~B. Mu\~{n}oz, Y.~Ali-Ha{\"\i}moud, M.~Kamionkowski,
  E.~D. Kovetz et~al., \emph{{Did LIGO detect dark matter?}},
  \href{https://doi.org/10.1103/PhysRevLett.116.201301}{\emph{Phys. Rev. Lett.}
  {\bfseries 116} (2016) 201301}
  [\href{https://arxiv.org/abs/1603.00464}{{\ttfamily 1603.00464}}].

\bibitem{2017JPhCS.840a2032G}
J.~{Garc{\'{\i}}a-Bellido}, \emph{{Massive Primordial Black Holes as Dark
  Matter and their detection with Gravitational Waves}},  in \emph{Journal of
  Physics Conference Series}, vol.~840 of \emph{Journal of Physics Conference
  Series}, p.~012032, May, 2017,
  \href{https://arxiv.org/abs/1702.08275}{{\ttfamily 1702.08275}},
  \href{https://doi.org/10.1088/1742-6596/840/1/012032}{DOI}.

\bibitem{1992PhRvD..46..645B}
J.~D. {Barrow}, E.~J. {Copeland} and A.~R. {Liddle}, \emph{{The cosmology of
  black hole relics}},
  \href{https://doi.org/10.1103/PhysRevD.46.645}{\emph{\prd} {\bfseries 46}
  (1992) 645}.

\bibitem{mack2008primordial}
K.~J. {Mack} and D.~H. {Wesley}, \emph{{Primordial black holes in the Dark
  Ages: Observational prospects for future 21cm surveys}}, {\emph{arXiv
  e-prints} (2008) arXiv:0805.1531}
  [\href{https://arxiv.org/abs/0805.1531}{{\ttfamily 0805.1531}}].

\bibitem{2010PhRvD..81j4019C}
B.~J. {Carr}, K.~{Kohri}, Y.~{Sendouda} and J.~{Yokoyama}, \emph{{New
  cosmological constraints on primordial black holes}},
  \href{https://doi.org/10.1103/PhysRevD.81.104019}{\emph{\prd} {\bfseries 81}
  (2010) 104019} [\href{https://arxiv.org/abs/0912.5297}{{\ttfamily
  0912.5297}}].

\bibitem{Boyle:2005se}
L.~A. Boyle and P.~J. Steinhardt, \emph{{Probing the early universe with
  inflationary gravitational waves}},
  \href{https://doi.org/10.1103/PhysRevD.77.063504}{\emph{Phys. Rev.}
  {\bfseries D77} (2008) 063504}
  [\href{https://arxiv.org/abs/astro-ph/0512014}{{\ttfamily
  astro-ph/0512014}}].

\bibitem{Malquarti:2003ia}
M.~Malquarti, S.~M. Leach and A.~R. Liddle, \emph{{From the production of
  primordial perturbations to the end of inflation}},
  \href{https://doi.org/10.1103/PhysRevD.69.063505}{\emph{Phys. Rev.}
  {\bfseries D69} (2004) 063505}
  [\href{https://arxiv.org/abs/astro-ph/0310498}{{\ttfamily
  astro-ph/0310498}}].

\bibitem{Easther:2006tv}
R.~Easther and H.~Peiris, \emph{{Implications of a Running Spectral Index for
  Slow Roll Inflation}},
  \href{https://doi.org/10.1088/1475-7516/2006/09/010}{\emph{\jcap} {\bfseries
  0609} (2006) 010} [\href{https://arxiv.org/abs/astro-ph/0604214}{{\ttfamily
  astro-ph/0604214}}].

\bibitem{Easther:2011yq}
R.~Easther and H.~V. Peiris, \emph{{Bayesian Analysis of Inflation II: Model
  Selection and Constraints on Reheating}},
  \href{https://doi.org/10.1103/PhysRevD.85.103533}{\emph{Phys. Rev.}
  {\bfseries D85} (2012) 103533}
  [\href{https://arxiv.org/abs/1112.0326}{{\ttfamily 1112.0326}}].

\bibitem{Page:1976df}
D.~N. Page, \emph{{Particle Emission Rates from a Black Hole: Massless
  Particles from an Uncharged, Nonrotating Hole}},
  \href{https://doi.org/10.1103/PhysRevD.13.198}{\emph{Phys. Rev.} {\bfseries
  D13} (1976) 198}.

\bibitem{1994caty.conf..132L}
J.~E. {Lidsey}, B.~J. {Carr} and J.~H. {Gilbert}, \emph{{Black Holes From Blue
  Spectra}},  in \emph{CMB Anisotropies Two Years after COBE: Observations,
  Theory and the Future} (L.~M. {Krauss}, ed.), p.~132, 1994,
  \href{https://arxiv.org/abs/astro-ph/9406028}{{\ttfamily astro-ph/9406028}}.

\bibitem{ivanov1998nonlinear}
P.~{Ivanov}, \emph{{Nonlinear metric perturbations and production of primordial
  black holes}}, \href{https://doi.org/10.1103/PhysRevD.57.7145}{\emph{Phys.
  Rev. D} {\bfseries 57} (1998) 7145}
  [\href{https://arxiv.org/abs/astro-ph/9708224}{{\ttfamily
  astro-ph/9708224}}].

\bibitem{2000astro.ph..3027C}
B.~J. {Carr} and C.~A. {Goymer}, \emph{{Primordial black holes and
  gravitational memory}}, {\emph{ArXiv Astrophysics e-prints} (2000) }
  [\href{https://arxiv.org/abs/astro-ph/0003027}{{\ttfamily
  astro-ph/0003027}}].

\bibitem{2007arXiv0708.3875H}
J.~C. {Hidalgo}, \emph{{The effect of non-Gaussian curvature perturbations on
  the formation of primordial black holes}}, {\emph{ArXiv e-prints} (2007) }
  [\href{https://arxiv.org/abs/0708.3875}{{\ttfamily 0708.3875}}].

\bibitem{carr2017primordial}
B.~{Carr}, T.~{Tenkanen} and V.~{Vaskonen}, \emph{{Primordial black holes from
  inflaton and spectator field perturbations in a matter-dominated era}},
  \href{https://doi.org/10.1103/PhysRevD.96.063507}{\emph{\prd} {\bfseries 96}
  (2017) 063507} [\href{https://arxiv.org/abs/1706.03746}{{\ttfamily
  1706.03746}}].

\bibitem{carr2018primordial}
B.~{Carr}, K.~{Dimopoulos}, C.~{Owen} and T.~{Tenkanen}, \emph{{Primordial
  black hole formation during slow reheating after inflation}},
  \href{https://doi.org/10.1103/PhysRevD.97.123535}{\emph{\prd} {\bfseries 97}
  (2018) 123535} [\href{https://arxiv.org/abs/1804.08639}{{\ttfamily
  1804.08639}}].

\bibitem{2000PhRvD..61b3501K}
K.~{Kohri} and J.~{Yokoyama}, \emph{{Primordial black holes and primordial
  nucleosynthesis: Effects of hadron injection from low mass holes}},
  \href{https://doi.org/10.1103/PhysRevD.61.023501}{\emph{\prd} {\bfseries 61}
  (2000) 023501} [\href{https://arxiv.org/abs/astro-ph/9908160}{{\ttfamily
  astro-ph/9908160}}].

\bibitem{Nayak:2009wk}
B.~Nayak and L.~P. Singh, \emph{{Accretion, Primordial Black Holes and Standard
  Cosmology}}, \href{https://doi.org/10.1007/s12043-011-0002-x}{\emph{Pramana}
  {\bfseries 76} (2011) 173} [\href{https://arxiv.org/abs/0905.3243}{{\ttfamily
  0905.3243}}].

\bibitem{sasaki2016primordial}
M.~{Sasaki}, T.~{Suyama}, T.~{Tanaka} and S.~{Yokoyama}, \emph{{Primordial
  Black Hole Scenario for the Gravitational-Wave Event GW150914}},
  \href{https://doi.org/10.1103/PhysRevLett.117.061101}{\emph{\prl} {\bfseries
  117} (2016) 061101} [\href{https://arxiv.org/abs/1603.08338}{{\ttfamily
  1603.08338}}].

\bibitem{inomata2017inflationary}
K.~{Inomata}, M.~{Kawasaki}, K.~{Mukaida}, Y.~{Tada} and T.~T. {Yanagida},
  \emph{{Inflationary primordial black holes for the LIGO gravitational wave
  events and pulsar timing array experiments}},
  \href{https://doi.org/10.1103/PhysRevD.95.123510}{\emph{\prd} {\bfseries 95}
  (2017) 123510} [\href{https://arxiv.org/abs/1611.06130}{{\ttfamily
  1611.06130}}].

\bibitem{inayoshi2016gravitational}
K.~{Inayoshi}, K.~{Kashiyama}, E.~{Visbal} and Z.~{Haiman},
  \emph{{Gravitational wave background from Population III binary black holes
  consistent with cosmic reionization}},
  \href{https://doi.org/10.1093/mnras/stw1431}{\emph{\mnras} {\bfseries 461}
  (2016) 2722} [\href{https://arxiv.org/abs/1603.06921}{{\ttfamily
  1603.06921}}].

\bibitem{ali2017merger}
Y.~{Ali-Ha{\"\i}moud}, E.~D. {Kovetz} and M.~{Kamionkowski}, \emph{{Merger rate
  of primordial black-hole binaries}},
  \href{https://doi.org/10.1103/PhysRevD.96.123523}{\emph{\prd} {\bfseries 96}
  (2017) 123523} [\href{https://arxiv.org/abs/1709.06576}{{\ttfamily
  1709.06576}}].

\bibitem{chen2018merger}
Z.-C. {Chen} and Q.-G. {Huang}, \emph{{Merger Rate Distribution of Primordial
  Black Hole Binaries}},
  \href{https://doi.org/10.3847/1538-4357/aad6e2}{\emph{\apj} {\bfseries 864}
  (2018) 61} [\href{https://arxiv.org/abs/1801.10327}{{\ttfamily 1801.10327}}].

\bibitem{nakamura1997gravitational}
T.~{Nakamura}, M.~{Sasaki}, T.~{Tanaka} and K.~S. {Thorne},
  \emph{{Gravitational Waves from Coalescing Black Hole MACHO Binaries}},
  \href{https://doi.org/10.1086/310886}{\emph{\apj} {\bfseries 487} (1997)
  L139} [\href{https://arxiv.org/abs/astro-ph/9708060}{{\ttfamily
  astro-ph/9708060}}].

\bibitem{peters1963gravitational}
P.~C. {Peters} and J.~{Mathews}, \emph{{Gravitational Radiation from Point
  Masses in a Keplerian Orbit}},
  \href{https://doi.org/10.1103/PhysRev.131.435}{\emph{Physical Review}
  {\bfseries 131} (1963) 435}.

\bibitem{peters1964gravitational}
P.~C. {Peters}, \emph{{Gravitational Radiation and the Motion of Two Point
  Masses}}, \href{https://doi.org/10.1103/PhysRev.136.B1224}{\emph{Physical
  Review} {\bfseries 136} (1964) 1224}.

\bibitem{2019JCAP...02..018R}
M.~{Raidal}, C.~{Spethmann}, V.~{Vaskonen} and H.~{Veerm{\"a}e},
  \emph{{Formation and evolution of primordial black hole binaries in the early
  universe}}, \href{https://doi.org/10.1088/1475-7516/2019/02/018}{\emph{\jcap}
  {\bfseries 2} (2019) 018} [\href{https://arxiv.org/abs/1812.01930}{{\ttfamily
  1812.01930}}].

\bibitem{1998PhRvD..58f3003I}
K.~{Ioka}, T.~{Chiba}, T.~{Tanaka} and T.~{Nakamura}, \emph{{Black hole binary
  formation in the expanding universe: Three body problem approximation}},
  \href{https://doi.org/10.1103/PhysRevD.58.063003}{\emph{\prd} {\bfseries 58}
  (1998) 063003} [\href{https://arxiv.org/abs/astro-ph/9807018}{{\ttfamily
  astro-ph/9807018}}].

\bibitem{2001astro.ph..8028P}
E.~S. {Phinney}, \emph{{A Practical Theorem on Gravitational Wave
  Backgrounds}}, {\emph{ArXiv Astrophysics e-prints} (2001) }
  [\href{https://arxiv.org/abs/astro-ph/0108028}{{\ttfamily
  astro-ph/0108028}}].

\bibitem{2016PhRvL.116m1102A}
B.~P. {Abbott}, R.~{Abbott}, T.~D. {Abbott}, M.~R. {Abernathy}, F.~{Acernese},
  K.~{Ackley} et~al., \emph{{GW150914: Implications for the Stochastic
  Gravitational-Wave Background from Binary Black Holes}},
  \href{https://doi.org/10.1103/PhysRevLett.116.131102}{\emph{Physical Review
  Letters} {\bfseries 116} (2016) 131102}
  [\href{https://arxiv.org/abs/1602.03847}{{\ttfamily 1602.03847}}].

\bibitem{easther2006stochastic}
R.~{Easther} and E.~A. {Lim}, \emph{{Stochastic gravitational wave production
  after inflation}},
  \href{https://doi.org/10.1088/1475-7516/2006/04/010}{\emph{Journal of
  Cosmology and Astro-Particle Physics} {\bfseries 2006} (2006) 010}
  [\href{https://arxiv.org/abs/astro-ph/0601617}{{\ttfamily
  astro-ph/0601617}}].

\bibitem{price2008stochastic}
L.~R. {Price} and X.~{Siemens}, \emph{{Stochastic backgrounds of gravitational
  waves from cosmological sources: Techniques and applications to preheating}},
  \href{https://doi.org/10.1103/PhysRevD.78.063541}{\emph{Phys. Rev. D}
  {\bfseries 78} (2008) 063541}
  [\href{https://arxiv.org/abs/0805.3570}{{\ttfamily 0805.3570}}].

\bibitem{husdal2016effective}
L.~{Husdal}, \emph{{On Effective Degrees of Freedom in the Early Universe}},
  \href{https://doi.org/10.3390/galaxies4040078}{\emph{Galaxies} {\bfseries 4}
  (2016) 78} [\href{https://arxiv.org/abs/1609.04979}{{\ttfamily 1609.04979}}].

\bibitem{2006PhRvL..97b1301S}
T.~L. {Smith}, E.~{Pierpaoli} and M.~{Kamionkowski}, \emph{{New Cosmic
  Microwave Background Constraint to Primordial Gravitational Waves}},
  \href{https://doi.org/10.1103/PhysRevLett.97.021301}{\emph{Physical Review
  Letters} {\bfseries 97} (2006) 021301}
  [\href{https://arxiv.org/abs/astro-ph/0603144}{{\ttfamily
  astro-ph/0603144}}].

\bibitem{2017JCAP...09..037R}
M.~{Raidal}, V.~{Vaskonen} and H.~{Veerm{\"a}e}, \emph{{Gravitational waves
  from primordial black hole mergers}},
  \href{https://doi.org/10.1088/1475-7516/2017/09/037}{\emph{\jcap} {\bfseries
  9} (2017) 037} [\href{https://arxiv.org/abs/1707.01480}{{\ttfamily
  1707.01480}}].

\bibitem{chisholm2006clustering}
J.~R. {Chisholm}, \emph{{Clustering of primordial black holes: Basic results}},
  \href{https://doi.org/10.1103/PhysRevD.73.083504}{\emph{\prd} {\bfseries 73}
  (2006) 083504} [\href{https://arxiv.org/abs/astro-ph/0509141}{{\ttfamily
  astro-ph/0509141}}].

\bibitem{chisholm2011clustering}
J.~R. {Chisholm}, \emph{{Clustering of primordial black holes. II. Evolution of
  bound systems}},
  \href{https://doi.org/10.1103/PhysRevD.84.124031}{\emph{\prd} {\bfseries 84}
  (2011) 124031} [\href{https://arxiv.org/abs/1110.4402}{{\ttfamily
  1110.4402}}].

\bibitem{2018arXiv180505912A}
Y.~{Ali-Ha{\"i}moud}, \emph{{Primordial black holes are initially not clustered
  beyond Poisson on small scales}}, {\emph{ArXiv e-prints} (2018) }
  [\href{https://arxiv.org/abs/1805.05912}{{\ttfamily 1805.05912}}].

\bibitem{desjacques2018spatial}
V.~{Desjacques} and A.~{Riotto}, \emph{{The Spatial Clustering of Primordial
  Black Holes}}, {\emph{arXiv e-prints} (2018) arXiv:1806.10414}
  [\href{https://arxiv.org/abs/1806.10414}{{\ttfamily 1806.10414}}].

\bibitem{ballesteros2018merger}
G.~{Ballesteros}, P.~D. {Serpico} and M.~{Taoso}, \emph{{On the merger rate of
  primordial black holes: effects of nearest neighbours distribution and
  clustering}},
  \href{https://doi.org/10.1088/1475-7516/2018/10/043}{\emph{Journal of
  Cosmology and Astro-Particle Physics} {\bfseries 2018} (2018) 043}
  [\href{https://arxiv.org/abs/1807.02084}{{\ttfamily 1807.02084}}].

\end{thebibliography}\endgroup

%\begin{thebibliography}{99}

%\bibitem{a}
%Author, \emph{Title}, \emph{J. Abbrev.} {\bf vol} (year) pg.

%\bibitem{b}
%Author, \emph{Title},
%arxiv:1234.5678.

%\bibitem{c}
%Author, \emph{Title},
%Publisher (year).

% Please avoid comments such as "For a review'', "For some examples",
% "and references therein" or move them in the text. In general,
% please leave only references in the bibliography and move all
% accessory text in footnotes.

% Also, please have only one work for each \bibitem.

%\end{thebibliography}
\end{document}